\documentstyle[11pt,newpasp,twoside,epsf]{article}
\def\edcomment#1{\iffalse\marginpar{\raggedright\sl#1\/}\else\relax\fi}
\reversemarginpar

\begin{document}
\title{BVRI Surface Photometry of (S+S) Binary Galaxies}
\author{Hector Hern\'andez Toledo}
\affil{IA-UNAM - Distrito Federal, Mexico}
\author{Iv\^anio Puerari}
\affil{INAOE - Tonantzintla, Mexico}

\begin{abstract}
We report the first results of our multicolour broad-band 
BVRI photometry for a statistically well defined sample 
of interacting pairs, drawn from the Catalogue of 
Isolated Pairs of Galaxies in the Northern Hemisphere 
Karachentsev (1972). At present, the details of photometric
structure of disk-disk (S+S) galaxies are just 
beginning to be studied because of their intrinsic 
difficulties. We take advantage of the statistical
properties of this well defined sample of interacting
galaxies to try to isolate the main structural
components of galaxies. Due to their environmental 
simplicity (compared to groups and clusters), the sample 
provides an unique opportunity to peform a photometric study 
of galaxies and their structural components in a 
non-equilibrium configuration. In this contribution, the 
first results of a deep multicolour (BVRI) photometric 
study of a sample of 45 (S+S) isolated pairs of galaxies 
is presented. Our attention is focused on the morphology 
and the global photometric properties (integrated magnitudes, 
colours and mean surface brightness profiles).
\end{abstract}

\section{Introduction}
Most statistical investigations of the properties of 
interacting galaxies focus on the most spectacular cases and 
thus they are affected by strong observational biases. The
samples have often been selected on basis of strongly peculiar
morphologies and/or high surface brightness. In this way, systems 
with a highly stimulated interstellar medium and increased star
formation rates have been favoured. However, some investigations
(eg., Kennicutt 1998; M\'arquez \& Moles 1996) have 
noted the problem and have made a special effort
to deal with homogeneous samples of interacting galaxies coupled
with well-matched comparison objects (isolated undisturbed
galaxies). These studies provide us with a more clear understanding about 
the connection between gravitational interactions, star 
formation and photometric properties in the disks of the 
intervening galaxies.

We have initiated an observational program to study the 
photometrical properties of a sample of disk--disk (S+S) 
systems with a wide range of separations, morphological 
distortions and presumably different stages of 
interactions. The use of a well defined isolation criterion in the 
selected pairs renders our study as ``clean'' as possible 
in the sense that only intrinsic properties of the individual 
galaxies and the effects of their mutual interactions should 
affect the observed photometrical properties.

In Figure 1, we present the mean surface brightness and
colour profiles for the galaxy pair KPG64. This is only an
example of the data we can obtain from our images. The
morphologies and global photometric properties of
the whole sample are being analysed for further publication
(Hern\'andez Toledo \& Puerari, in preparation).

\begin{figure}
\vskip8truecm
\includegraphics{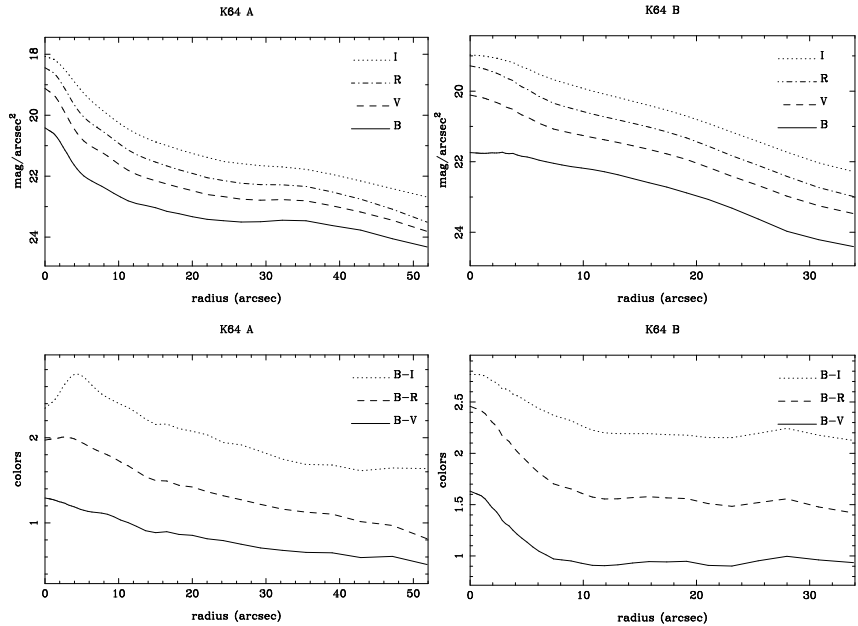}
\caption{Mean surface brightness and colour profiles for the
galaxy pair KPG64.}
\end{figure}

The observations discussed here are part of our 
broad-band BVRI photometric program on binary systems 
which is being carried out at the two Mexican Optical
Observatories: The Observatorio Astron\'omico Guillermo 
Haro in Cananea, Sonora, Mexico and the Observatorio
Astron\'omico Nacional in San Pedro Martir, Baja California,
Mexico.
 
\acknowledgments
HHT and IP were fortunate enough to have
participated in this meeting and want to
thank the organizers.

\end{document}